\documentclass[dvipdfmx]{elsarticle}

\biboptions{sort&compress}
\usepackage{graphicx}
\usepackage{amsmath,amsthm,amssymb}
\usepackage{bm}

\newcommand{\V}{\bm}

\newlength{\figurewidth}
\figurewidth = \textwidth
\advance \figurewidth by -\columnsep
\divide  \figurewidth by 2

\journal{Nuclear Instruments and Methods in Physics Research A }

\begin{document}

\begin{frontmatter}
  
  \title{Bin Mode Estimation Methods for Compton Camera Imaging}

  \author[ism,gaus]{S.~Ikeda\corref{cor}}
  \cortext[cor]{Corresponding author}
  \ead{shiro@ism.ac.jp}

  \author[jaxa]{H.~Odaka}
  \author[hirodai]{M.~Uemura}
  \author[jaxa]{T.~Takahashi}
  \author[jaxa]{S.~Watanabe}
  \author[jaxa]{S.~Takeda}

  \address[ism]{The Institute of Statistical Mathematics, Tachikawa,
    Tokyo, 190-8562, Japan}
  \address[gaus]{The Graduate University for Advanced Studies,
    Kanagawa, 240-1542, Japan}
  \address[jaxa]{Japan Aerospace Exploration Agency, Sagamihara,
    Kanagawa, 252-5210, Japan}
  \address[hirodai]{Hiroshima University, Higashi-hiroshima,
    Hiroshima, 739-8511, Japan}

\begin{abstract}
  We study the image reconstruction problem of a Compton camera which
  consists of semiconductor detectors. The image reconstruction is
  formulated as a statistical estimation problem. We employ a bin-mode
  estimation (BME) and extend an existing framework to a Compton
  camera with multiple scatterers and absorbers. Two estimation
  algorithms are proposed: an accelerated EM algorithm for the maximum
  likelihood estimation (MLE) and a modified EM algorithm for the
  maximum a posteriori (MAP) estimation. Numerical simulations
  demonstrate the potential of the proposed methods.
\end{abstract}


\end{frontmatter}

\section{Introduction}
\label{Introduction}

Compton camera imaging is a promising method to visualize gamma-ray
sources from $100\mathrm{keV}$ to several $\mathrm{MeV}$. It can be
applied in many fields, such as nuclear medicine, visualization of
radioactive substances on the ground, and gamma-ray astronomy
\cite{Bloemem.etal.1994.ajs,Schoenfelder.etal.1996aass,Knoedlseder_etal.1999.aa,LeBlanc.eta.1998.ieeens,Takahashi.etal.2003spie,Kanbach.etal.2004.nar,Boggs.2006.nar,jaxa2012}.
High-resolution Compton cameras utilizing silicon (Si) and cadmium
telluride (CdTe) semiconductor detectors
\cite{Watanabe.etal.2005.ieeens,Takeda_phd,Takeda.etal2009.ieeens,Odaka.etal.2012.nima}
will be installed in the next-generation X-ray observatory ASTRO-H
\cite{Takahashi.etal.2010.spie,Tajima.etal.2010.spie}, which is
scheduled for launch in 2015.

However, imaging is not straightforward because only a small portion
of photons are absorbed after Compton scattering and the direction of
arrival of each photon is not known directly. From a single event, the
scattering angle of the photon is computed using the energies of the
recoil electron and the scattered photon. After collecting these
observations, some type of information processing is needed to
reconstruct the image. Since photon detection is a stochastic process,
the image reconstruction problem can be formulated as a statistical
estimation problem
\cite{Parra2000ieeetns,HirasawaTomitani2002pmb,XuHe2006ieeens}. In
this study, we follow the framework developed for COMPTEL
\cite{Bloemem.etal.1994.ajs,Knoedlseder_etal.1999.aa,Bandstra_etal.2011.ap},
and employ the bin-mode estimation (BME) method. Although the number
of bins can be large, in astronomy applications where the distances
from the gamma-ray sources are large, the number of bins is
significantly reduced and the BME method is applied effectively.

One of the popular estimation methods in statistics is the maximum
likelihood estimation (MLE). COMPTEL also employed the MLE. In order
to compute the MLE of the Compton camera imaging, a natural approach
is to use the expectation-maximization (EM) algorithm
\cite{Dempster.etal.1977.jrssB}. However, the convergence speed of the
EM algorithm is not fast in general, and a line search algorithm was
combined with the EM algorithm in COMPTEL. In this work, we propose
two different approaches to speed up the convergence of the EM
algorithm: one is a different acceleration method to compute the MLE
by approximating the Fisher's scoring method
\cite{Ikeda2000scj,McLachlanKrishnan2008john} and the other is to use
the maximum a posteriori (MAP) estimation instead of the MLE. The MAP
estimation is another popular estimation method in Bayesian statistics
\cite{BernardoSmith.1994.john,Bishop2006springer}. We employed the
Dirichlet distribution as the prior, and a modified EM algorithm is
used to compute the MAP estimate. These two proposed BME methods are
tested through numerical simulations.

The rest of the paper is organized as follows. 
Section \ref{sec:Compton camera imaging} explains the Compton camera
system, while section \ref{sec:BME methods} overviews BME methods for
image reconstruction. Section \ref{sec:Numerical Results} shows
numerical results with some discussions, and section
\ref{sec:Conclusion} concludes the paper.

\section{Compton camera imaging}
\label{sec:Compton camera imaging}

\subsection{Compton camera system}
\label{subsec:Compton camera system}

\begin{figure}[hbtp]
  \centering
  \includegraphics[width=\figurewidth]{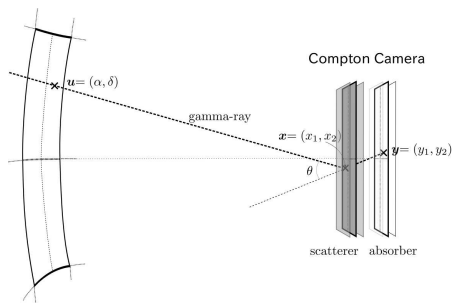}
  \caption{Compton camera.}
  \label{fig:comptoncamera}
\end{figure}
Figure \ref{fig:comptoncamera} schematically depicts the Compton
camera system. Our target system has multiple scatterers and
absorbers. A photon is detected when it is scattered at one of the
scatterers and absorbed by one of the absorbers. Even if a number of
photons arrive, only a little portion of them are detected after
Compton scattering.

Suppose a photon from $\V{u}=(\alpha,\delta)$ (in equatorial
coordinates) is detected. It is initially scattered by one of the
scatterers at $\V{x}=(x_1,x_2)$ and then absorbed by one of the
absorbers at $\V{y}=(y_1,y_2)$. We denote the distance between the
scatterer and the absorber as $d$. 

Let $E_1$ and $E_2$ be the energies of the recoil electron and the
scattered photon, respectively. The scattering angle $\theta$ is
denoted as
\begin{align}
  \label{eq:costh}
  \cos\theta =1-m_ec^2\left(\frac{1}{E_2}-\frac{1}{E_1+E_2}\right),
\end{align}
where $m_e$ and $c$ are the mass of an electron and the speed of
light, respectively. The goal of Compton camera imaging is to
reconstruct the gamma-ray intensity map on the celestial sphere from
collected information. Difficulties arise because only the scattering
angle $\theta$ is computed from a single event.

In this paper, we assume the distance from each gamma-ray source to
the Compton camera system is sufficiently large. This assumption is
valid for astronomy applications. Under this assumption, only the
direction of arrival is important, therefore, not $\V{x}$ and $\V{y}$,
but only the relative position normalized by the distance $d$, i.e.,
$\V{w}=(\V{x}-\V{y})/d$, is considered. The information of each
detected photon is summarized into $\V{v} =
(\V{w},\cos\theta)$. Furthermore $\V{u}=(\alpha,\delta)$ and $\V{v}$
are assumed to be quantized into bins. Below, $\V{u}$ and $\V{v}$ are
bin indices.

\subsection{Compton camera measurement process}
\label{subsec:Compton camera measurement process}

We first introduce the model of the measurement process
\cite{Wilderman.etal.2001.ieeens}. When a
gamma-ray photon from $\V{u}$ is detected, it falls into a bin $\V{v}$
with some probability. Let $\lambda(\V{u})$ be the intensity of the
gamma rays at pixel $\V{u}$ of the celestial sphere and $Y(\V{v})$ be
the number of photons detected at bin $\V{v}$ for a time interval. The
distribution of $Y(\V{v})$ follows a Poisson distribution as
\begin{align}
  \label{eq:poisson}
  Y(\V{v})\sim \mathrm{Poisson} 
  \left(\sum_{\V{u}}t({\V{v}},\V{u})~\lambda(\V{u})\ \right).
\end{align}
where $t({\V{v}},\V{u})$ is the probability that a photon from $\V{u}$
is absorbed at $\V{v}$.
We modify the above framework in this work. Let us re-parameterize
$t({\V{v}},\V{u})$ and $\lambda(\V{u})$ and define $q(\V{v})$ as
follows
\begin{align}
  \begin{split}
  p({\V{v}}\,|\,\V{u})&=\frac{t({\V{v}},\V{u})}{s(\V{u})}
  ~~~{\mathrm{where}}~~~
  s(\V{u}) = \sum_{\V{v}} t({\V{v}},\V{u}),\\
  \rho(\V{u})&= \frac{\lambda(\V{u})s(\V{u})}{\lambda}
  ~~~{\mathrm{where}}~~~\lambda =
  \sum_{\V{u}}\lambda(\V{u})s(\V{u})\\
  q(\V{v})&=\sum_{\V{u}}p({\V{v}}\,|\,\V{u})\,\rho(\V{u}),
  \end{split}
\end{align}
where $s(\V{u})$ is the probability that a photon from pixel $\V{u}$
is absorbed by the camera. All $\rho(\V{u})$, $p(\V{v}\,|\,\V{u})$,
and $q(\V{v})$ are multinomial distributions (i.e., they are
non-negative and each distribution sums up to one) conditional to the
events that a photon is absorbed by the camera. More precisely,
$\rho(\V{u})$ is the probability that an absorbed photon arrives from
$\V{u}$, $p({\V{v}}\,|\,\V{u})$ is the probability that an absorbed
photon from $\V{u}$ is detected at $\V{v}$, and $q(\V{v})$ is the
probability that an absorbed photon is detected at $\V{v}$. After
observing many photons, we collect information of $q(\V{v})$. If
$p(\V{v}\,|\,\V{u})$ is known, it is possible to estimate
$\rho(\V{u})$.

The goal of Compton camera imaging is to reconstruct the intensity map
$\lambda(\V{u})$. The distribution $\rho(\V{u})$ differs from
$\lambda(\V{u})$ because the photons which were not completely
absorbed are not included. However $\lambda(\V{u})$ can be easily
reconstructed as $\lambda(\V{u}) \propto {\rho(\V{u})}/{s(\V{u})}$
assuming $s(\V{u})$ is known. Thus, we set our goal to estimate
$\rho(\V{u})$ from absorbed photons.

In the next section, we first show how to prepare $p(\V{v}\,|\,\V{u})$
and $s(\V{u})$, then explain how to estimate $\rho(\V{u})$ from
absorbed photons.

\section{BME methods}
\label{sec:BME methods}

\subsection{Bin mode}
\label{subsec:Bin mode estimation}

Our formulation is based on the BME, where each $\V{v}$ and $\V{u}$ is
quantized into bins. The key is how to implement and prepare
$p(\V{v}\,|\,\V{u})$.

If $\V{x}$, $\V{y}$, $E_1$, and $E_2$ are quantized separately, the
number of the bins is too large and the BME is not feasible with
current computational hardware. However, under the assumption that the
distances from the gamma ray sources are large, it is sufficient to
quantify $(\V{w},\cos\theta)$, where $\V{w}=(\V{x}-\V{y})/d$ and
$\cos\theta$ are defined in eq.\,\eqref{eq:costh}. The number of bins
decreases and the BME becomes feasible without loss to precision.
In previous works of COMPTEL, the matrix size was further reduced
using geometrical symmetries
\cite{Bloemem.etal.1994.ajs,Knoedlseder_etal.1999.aa} but we do not
rely on symmetry because the installed camera may have an asymmetric
configuration.

The next challenge is how to prepare $p(\V{v}\,|\,\V{u})$ and
$s(\V{u})$.  Here we utilize a numerical method. A Compton camera
system was simulated and a lot of photons were randomly drawn
numerically using software such as Geant4. The results were
accumulated to compute $p(\V{v}\,|\,\V{u})$ and $s(\V{u})$. This
method is general and easily implemented.

There are other possible methods to compute $p(\V{v}\,|\,\V{u})$ and
$s(\V{u})$. One of them is to used the numerical integration based on
a physical model with the Klein-Nishina formula. Although the
integration is difficult, the resulting distribution is theoretically
accurate. Another method is to use a physical system. When a system is
built and tested under a physical environment, the collected data from
a set of well-designed experiments can be used to compute
$p(\V{v}\,|\,\V{u})$ and $s(\V{u})$.

\subsection{Maximum likelihood estimate}
\label{subsec:MLE and MAP}

Next, we discuss the estimation of $\rho(\V{u})$. 

Suppose $N$ photons are detected independently and $t$-th photon is
absorbed at $\V{v}_t$, $(t=1,\cdots,N)$. The log likelihood function
is defined as follows
\begin{align}
  \label{eq:log likelihood}
  L(\rho) = \sum_{t=1}^N \log\, q(\V{v}_t)= \sum_{t=1}^N \log \sum_{\V{u}}
  p(\V{v}_t\,|\,\V{u})~\rho(\V{u}).
\end{align}
Our goal is to estimate $\rho(\V{u})$ from $\{\V{v}_t\}$, where
$p(\V{v}\,|\,\V{u})$ is known. The maximum likelihood estimate (MLE)
of $\rho(\V{u})$ is defined as the distribution $\rho(\V{u})$ that
maximizes $L(\rho)$. We denote it as
$\hat{\rho}_{\mathrm{MLE}}(\V{u})$. The EM algorithm
\cite{Dempster.etal.1977.jrssB} is a simple algorithm to compute MLE
by alternately repeating the expectation (E-) step and maximization
(M-) step. Each step is defined as
\begin{align}
  \label{eq:e-step}
  &(\mbox{E-step})
    &&q^{(l)}(\V{v}) 
    = \sum_{\V{u}}p(\V{v}\,|\,\V{u})~\rho^{(l)}(\V{u}),\\
  \label{eq:m-step}
  &(\mbox{M-step})
    &&\rho^{(l+1)}(\V{u})
    = 
    \frac1N
    \sum_{t=1}^N
    \frac{p(\V{v}_t\,|\,\V{u})}{q^{(l)}(\V{v}_t)}~\rho^{(l)}(\V{u}),
\end{align}
where $l$ starts from $0$ and increases by 1 at each iteration. This
algorithm can be found in a literature
\cite{SheppVardi1982ieeemi}. The log-likelihood is non-decreasing at
each update, that is, $L(\rho^{(l+1)})\ge L(\rho^{(l)})$. When the
difference between $q^{(l+1)}(\V{v})$ and $q^{(l)}(\V{v})$ becomes
sufficiently small, the update is terminated. In order to measure the
difference, we employed the Kullback-Leibler (KL) divergence which is
defined as follows,
\begin{align*}
 KL(q^{(l+1)}, q^{(l)}) =
 \sum_{\V{v}} q^{(l+1)}(\V{v}) \log \frac{q^{(l+1)}(\V{v})}{q^{(l)}(\V{v})}.
\end{align*}
Note that the KL divergence is non-negative and becomes 0 if and only
if $q^{(l+1)}(\V{v})=q^{(l)}(\V{v})$ for all $\V{v}$. We stopped the
algorithm when $KL(q^{(l+1)}, q^{(l)})$ is less than $1.0\times
10^{-10}$. The reconstructed image is proportional to the intensity
map $\lambda(\V{u})$, which is recovered from
$\hat{\rho}_{\mathrm{MLE}}(\V{u})$ as
$\hat{\rho}_{\mathrm{MLE}}(\V{u})/s(\V{u})$.

\subsection{Acceleration of the EM algorithm}
\label{subsec:acceleration}

The convergence speed of the EM algorithm is generally slow. In this
work, we implemented an acceleration algorithm which approximates the
Fisher's scoring method
\cite{Ikeda2000scj,McLachlanKrishnan2008john}. The Fisher's method is
a second order method, similar to the Newton's method, and a faster
convergence is expected. The proposed acceleration method was combined
with a line search algorithm as in COMPTEL
\cite{Knoedlseder_etal.1999.aa}.

We show the outline of the acceleration algorithm
\cite{Ikeda2000scj,McLachlanKrishnan2008john}. When
$\rho^{(l+1)}(\V{u})$ is computed from $\rho^{(l)}(\V{u})$ with one
EM-step, $q^{(l+1)}(\V{v})$ is computed from
eq.\,\eqref{eq:e-step}. Then another EM-step is run from
$\rho^{(l)}(\V{u})$ where $q^{(l+1)}(\V{v})$ is the target
distribution. After this EM-step, a new parameter ${\rho'}(\V{u})$ is
obtained, which differs from $\rho^{(l+1)}(\V{u})$ and the following
parameter is computed
\begin{align}
  \rho^{new}(\V{u})
  \propto \frac{\rho^{(l+1)}(\V{u})^2}{\rho'(\V{u})}.
\end{align}
In most cases $L(\rho^{new}(\V{u}))$ is larger than
$L(\rho^{(l+1)}(\V{u}))$ and the EM algorithm is accelerated
\cite{Ikeda2000scj}. 

The convergence speeds of the proposed acceleration method and the
original EM algorithm are compared through numerical simulations in
section \ref{subsec:Computational time of algorithms}.

\subsection{Maximum a posteriori estimate}
\label{subsec:MAP}

Starting from a strictly positive initial distribution
$\rho^{(0)}(\V{u})>0$, every component of
$\hat{\rho}_{\mathrm{MLE}}(\V{u})$ is strictly positive by
definition\footnote{Some components become smaller than the numerical
  precision and are set to $0$.}. However, for astronomy applications,
it is natural to assume there are a lot of $0$ components. To estimate
such a sparse solution, we used the maximum a posteriori (MAP)
estimation, which is a common approach in Bayesian statistics
\cite{BernardoSmith.1994.john,Bishop2006springer}. 

Setting the prior of $\rho$ as $\pi(\rho)$, the MAP estimation
maximizes the posterior probability
$P(\rho,|\,\V{v}_1,\cdots,\V{v}_N)$, which is proportional to
$\pi(\rho)\prod_tq(\V{v}_t)$. That is,
\begin{align}
  \label{eq:posterior}
  \begin{split}
  \hat{\rho}_{\mathrm{MAP}}
  &= \arg\max_{\rho}\,
  \Bigl[\,
  \log \,P(\rho\,|\,\V{v}_1,\cdots,\V{v}_N)\,
  \Bigr]
  \\
  &= \arg\max_{\rho}\,
  \Bigl[\,
  \log \,\pi(\rho)+L(\rho)
  \Bigr],
  \end{split}
\end{align}
The Dirichlet distribution is the conjugate prior of a multinomial
distribution \cite{BernardoSmith.1994.john,Bishop2006springer} and is
often used as the prior for multinomial distributions. We use the
following symmetric Dirichlet distribution with a single parameter
$\beta$ as the prior of $\rho(\V{u})$,
\begin{align}
  \label{eq:dirichlet}
  \pi_{\beta}(\rho) = \frac{\Gamma(M \beta)}{\Gamma(\beta)^M}
  \prod_{\V{u}}\rho(\V{u})^{\beta-1}, 
\end{align}
where $M$ is the number of bins of $\V{u}$ and $\beta>0$.  With this
prior, eq.\,\eqref{eq:posterior} is rewritten as $
\hat{\rho}_{\mathrm{MAP}} = \arg\max_{\rho}\, \Bigl[\,
(\beta-1)\sum_{\V{u}} \log\, \rho(\V{u})+ L(\rho)\, \Bigr]$. Note
that as the number of the samples increases, the log likelihood
function defined in eq.\,\eqref{eq:log likelihood} increases, and the
influence of the prior becomes relatively small. In the limit $N\to
\infty$, the MAP estimate becomes identical to the MLE. 

The EM algorithm can be used to compute the MAP estimate by modifying
the M-step as
\begin{align}
  \label{eq:MAP EM}
  &(\mbox{M-step})
  &&\rho^{(l+1)}(\V{u})\propto\max\, 
  \biggl[0,~
  \Bigl(\beta-1 + \sum_{t=1}^N
  \frac{p(\V{v}_t\,|\,\V{u})}{q^{(l)}(\V{v}_t)}~\rho^{(l)}(\V{u})
  \Bigr)
  \biggr].
\end{align}
If $\beta=1$, the prior is uniform and the MAP estimation is identical
to the MLE. However, if $\beta<1$, $\hat{\rho}_{\mathrm{MAP}}(\V{u})$
has 0 components. The number of 0 components tends to increase as
$\beta$ decreases. Although the prior in eq.\,\eqref{eq:dirichlet} is
improper for $\beta\le 0$, the updating rule in eq.\,\eqref{eq:MAP EM}
is still valid. In the rest of the paper, we set $\beta=0$. It should
be noted that the convergence speed of the above modified EM algorithm
is generally faster than the original EM algorithm. The MAP estimates
are computed quickly if many components of $\rho(\V{u})$ are 0. We
show some numerical results in the next section.

We proposed the MAP estimation in order to promote some components to
be 0, in other word, to have a sparse solution. One concern is that
some weak, possibly distributed gamma-ray sources may be neglected by
the MAP estimation. When the number of received photons is small, this
may be true but the influence of the prior becomes smaller as the
number of absorbed photons increases. If a sufficiently large number
of photons are received, the MAP reconstruction is almost identical to
the MLE reconstruction and any weak sources would not be neglected.

\section{Numerical Results}
\label{sec:Numerical Results}

\subsection{Simulated Compton camera}

\begin{figure}[htbp]
  \centering
  \includegraphics[width = 0.7\figurewidth]{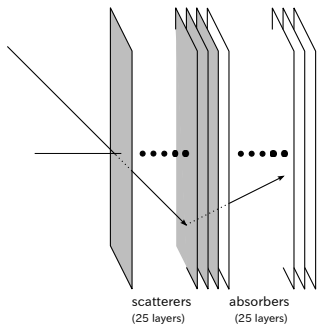}
  \caption{Simulated Compton camera with 25 layered scatterers (Si)
    and absorbers (CdTe). Each detector is 1.0{$\,\mathrm{mm}$} thick
    with a 2{$\,\mathrm{mm}$} separation.}
  \label{fig:layered}
\end{figure}

A Compton camera, which consists of 25 Si pixel detectors as the
scatterers and 25 CdTe pixel detectors as the absorbers, was simulated
for the numerical experiments (fig.\,\ref{fig:layered}). The size of
each detector was identical: 100{$\,\mathrm{mm}$} square and
1.0{$\,\mathrm{mm}$} thick. All of the detectors were stacked in
2{$\,\mathrm{mm}$} intervals. Each detector was pixellated into $50
\times 50$ channels, and was assumed to have an energy resolution of
$\sqrt{(2\mathrm{keV})^2 + E^2}$ (full width at half maximum, FWHM),
where $E$ denotes the energy deposited in the detector.

In order to perform Monte Carlo simulation,
a model of the Compton camera has been constructed.
We used our Compton camera response generator
\cite{Odaka.etal.2010.nima}, which was implemented using the Geant4
toolkit (Version 9.6.p01) \cite{Allison.etal.2006.ieeens} on a desktop
computer (Mac Pro 6-core Intel Xeon 3.33\,GHz). The simulations
accurately treated the Doppler broadening effect of Compton scattering
\cite{ZoglauerKanbach.2003.spie} due to the non-zero momentum of the
target electron, which degrades the angular resolutions of the Compton
camera. Using this simulation framework, we estimated the angular
resolution measure (ARM) of the simulated Compton camera at 2.6
degrees (FWHM) for a photon energy of 661.7\,keV. The ARM is defined
as the difference between the reconstructed scattering angle and the
actual source direction \cite{Takeda_phd,Takeda.etal2009.ieeens}.

In the current simulation, the range of $\alpha$ and $\delta$ were
restricted between $\pm 30^{\circ}$. It is possible to widen these
ranges, but the goal of this work is to build effective estimation
methods. Thus, we stayed within $\pm 30^{\circ}$. In a real system,
events from outside the field of view can be discarded by physical
shielding and/or background rejection techniques based on Compton
camera analysis prior to image reconstruction.

\begin{table}[hbtp]
  \caption{Design of bins of $\V{u}$ and $\V{v}$.}
  \label{tab:mesh}
  \centering
  \begin{tabular}[h]{c|cc|ccc}
    \hline
    & \multicolumn{2}{c|}{$\V{u}$[degree]} & 
    \multicolumn{3}{c}{$\V{v}$}\\
    & {$\alpha$} & {$\delta$} & 
    {\small $\displaystyle\frac{w_1}{\sqrt{\,|w_1|\,}}$} &
    {\small $\displaystyle\frac{w_2}{\sqrt{\,|w_2|\,}}$} &
    {\small $\cos\theta$}
    \\\hline
    min & $-30$ & $-30$ & $-2.0$ & $-2.0$ &
    $-0.56$\\
    max & $30$ & $30$ & $2.0$ & $2.0$ & $0.96$\\\hline
    $\#$ of bins & $47$ & $47$ & $21$ & $21$ & $32$\\\hline
  \end{tabular}
\end{table}
The designs of image and data space is summarized in
table\,\ref{tab:mesh}. Angles $\alpha$, $\delta$, and $\cos\theta$
were quantized into equally spaced bins. For relative positions $w_1$
and $w_2$, equal spacing is not preferable because more photons are
absorbed at the center than the boundaries. To prevent such unbalanced
measurements, the square roots of $w_1$ and $w_2$ were quantized into
equally spaced bins. The number of photons absorbed at each bin became
similar under this quantization. The total numbers of bins of $\V{u}$
and $\V{v}$ were $47\times 47 = 2209$ and $21\times 21 \times 32
=14,112$, respectively. Each image bin was set to half of the ARM
FWHM. The data space binning was also roughly optimized by evaluating
the angular broadening of a point source reconstruction.

\begin{figure}[htb]
  \centering 
  \includegraphics[width=1.2\figurewidth]{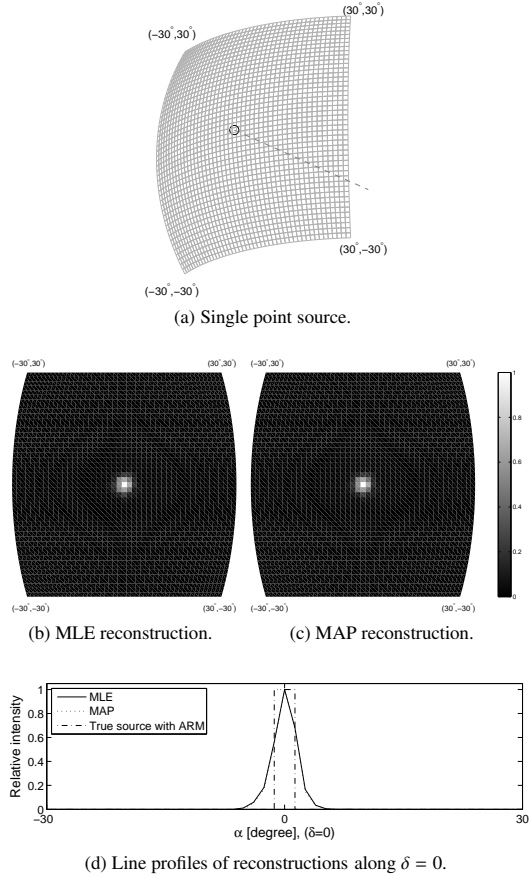}
  \caption{Gamma-ray source and estimation results.
    (a) Single point source. Circle indicates the position of a
    point source at $(0^{\circ},0^{\circ})$. Dashed line indicates the
    direction of the Compton camera.
    (b) MLE reconstruction,
    (c) MAP reconstruction (98 pixels are positive), and (d)
    line-profiles of images in (b) and (c) compared with the true
    source profile. Each image was normalized to have $1$ as its
    maximum.}
  \label{fig:fig0}
\end{figure}

In order to compute $p(\V{v}\,|\,\V{u})$ and $s(\V{u})$, a large
number of photons, $2.4\times 10^{10}$, were randomly generated and
recorded. Total computational time for this simulation was 270 hours
using an Apple Mac Pro 6-core Xeon 3.33 GHz (12 processes with
hyper-threading). The energy of a gamma-ray photon was set to
661.7\,keV ($^{137}\mathrm{Cs}$). The number of detected photons was
28,817,844, which was around 0.120$\,\%$ of the total photons. The
distributions $p(\V{v}\,|\,\V{u})$ and $s(\V{u})$ were computed with
these photon counts. It should be noted that the estimated
$p(\V{v}\,|\,\V{u})$ had some 0's, which could make the EM algorithm
unstable. We replaced $0$'s with a small constant, $1.0\times
10^{-10}$.

This simulation only considered $^{137}\mathrm{Cs}$. In order to apply
this method for different energy levels, $p_{E}(\V{v}\,|\,\V{u})$ and
$s_{E}(\V{u})$ must be prepared for each energy level $E$, and the
gamma ray intensities must be estimated separately depending on the
energy level $E = E_1+E_2$.

\subsection{Computational times for different algorithms}
\label{subsec:Computational time of algorithms}

We used a simple point source example in fig.\,\ref{fig:fig0} (a) to
test the estimation algorithms. The number of simulated photons was
30,000,000 and only 108,557 photons were absorbed. We reconstructed
the gamma-ray source from the detected photons.

Three algorithms were tested: the EM algorithm for the MLE, the
accelerated EM algorithm for the MLE, and the modified EM algorithm
for the MAP estimation. Table\,\ref{tab:comptime} shows the
computational times and iteration numbers for convergence.
\begin{table}[htb]
  \centering
  \caption{Computational times and iteration numbers fir the three
    algorithms implemented on a desktop computer (Intel Core i7-3770S,
    3.10\,GHz) with Matlab 2013a (Ubuntu Linux 13.04, x86-64).}
  \label{tab:comptime}
  \begin{tabular}[h]{lrr}
    \hline
    &time [s]& iterations\\
    \hline
    EM (MLE) & 370 & 1563\\
    accelerated EM (MLE) & 133 & 514\\
    modified EM (MAP) & 7.5 & 755\\
    \hline
  \end{tabular}
\end{table}

Figure\,\ref{fig:fig0} shows the results. The reconstructed images of
the two MLE algorithms are identical to fig.\,\ref{fig:fig0} (b), and
the MAP reconstruction is very similar (fig.\,\ref{fig:fig0}
(c)). However, the MAP reconstruction has only 98 positive pixels,
while 2209 ($=47\times 47$) pixels are positive for the MLE
reconstruction. The profiles of reconstructed images
(figs.\,\ref{fig:fig0} (b) and (c)) along the line $\delta=0$ are
shown in fig.\,\ref{fig:fig0} (d). It is difficult to distinguish
between MLE and MAP reconstructions. The vertical bar in
fig.\,\ref{fig:fig0} (d) indicates the location of the true source
point. The width of the bar is equal to the ARM. Proposed methods
provided good estimates of the true source location, and the positive
regions of the reconstructed images were similar to the ARM width.

We have shown that the accelerated EM algorithm converged faster than
the original EM algorithm. Moreover, the MAP estimation was very fast,
and converged to a reasonable sparse image.

\subsection{Image reconstruction}

\begin{figure}[htb]
  \centering 
  \includegraphics[width=1.2\figurewidth]{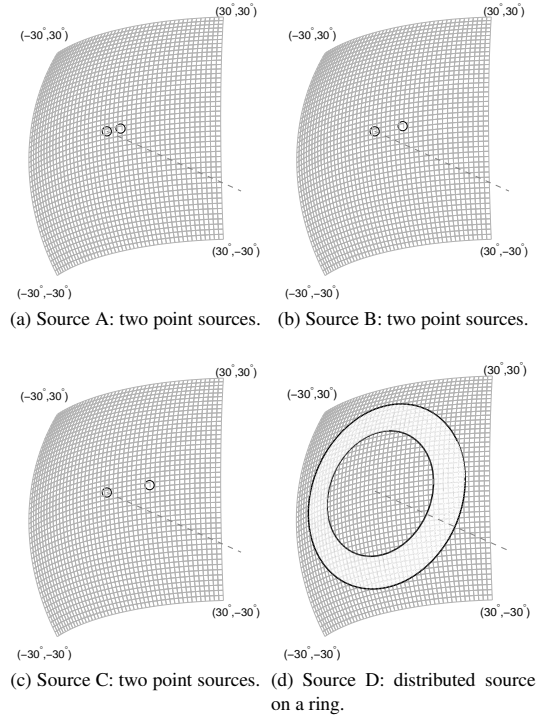}
  \caption{Gamma-ray sources used for numerical simulations:
    (a) source A: the circles indicate the positions of point sources at
    $(0^{\circ},0^{\circ})$ and $(4^{\circ},0^{\circ})$.
    (b) source B: two point sources at
    $(0^{\circ},0^{\circ})$ and $(8^{\circ},0^{\circ})$,
    (c) source C: two point sources at
    $(0^{\circ},0^{\circ})$ and $(12^{\circ},0^{\circ})$, 
    and (d) source D: distributed source on a ring, where the outer
    and inner circles were $16^\circ$ and $24^\circ$ apart from the
    center, respectively. Dashed lines indicate the direction of the
    Compton camera.}
  \label{fig:sources}
\end{figure}

We next prepared four types of gamma-ray sources for numerical
simulations. Sources A, B and C were two point sources
(figs.\,\ref{fig:sources}a, \ref{fig:sources}b and \ref{fig:sources}c,
respectively), and source D was a distributed source on a ring-shaped
region (fig.\,\ref{fig:sources}d).

\begin{figure}[htbp]
  \centering
  \includegraphics[width=1.2\figurewidth]{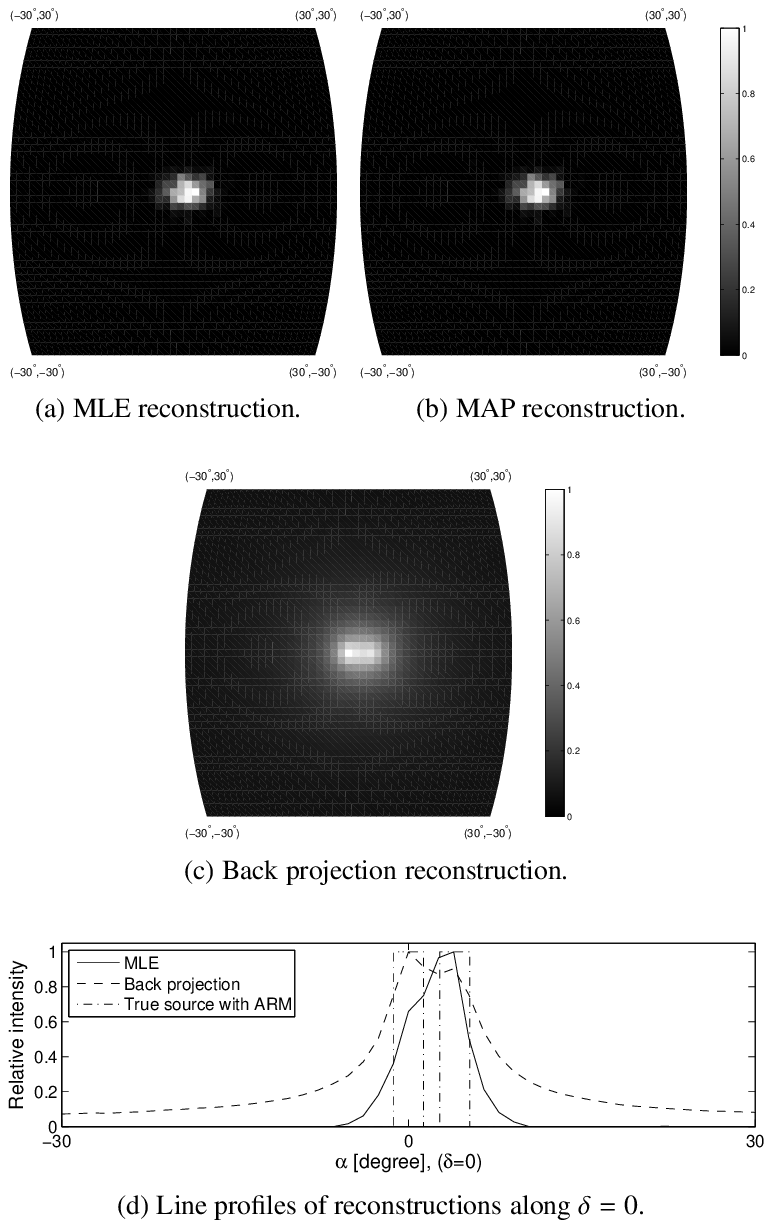}
  \caption{Reconstructed images for source A in
    fig.\,\ref{fig:sources}a. using 108,557 absorbed photons.
    (a) MLE reconstruction, 
    (b) MAP reconstruction (127 pixels are positive),
    (c) image reconstruction with back projection, and (d)
    line-profiles of images in (a) and (c) compared with the true
    source profile. Each image was normalized to have $1$ as its
    maximum.}
  \label{fig:result0}
\end{figure}

From each source model, a number of photons were randomly generated,
and images were reconstructed. We compared the results of the MLE
computed by the accelerated EM algorithm, the MAP estimation with the
Dirichlet prior distribution ($\beta=0$), and the back-projection.

Figure \ref{fig:result0} shows the reconstructed images for source
A. The reconstructed images of the BME methods have compact positive
regions but two point sources are not separated
(figs.\,\ref{fig:result0} (a), (b) and (d)). The back-projection
reconstruction is rather diffused, but it has two peaks corresponding
to the two point sources (figs.\,\ref{fig:result0} (c) and
(d)). Though the MLE and MAP reconstructions looks similar, each pixel
of the MLE reconstruction is strictly positive whereas the MAP
reconstruction has only 127 positive pixels.

\begin{figure}[htbp]
  \centering
  \includegraphics[width=1.2\figurewidth]{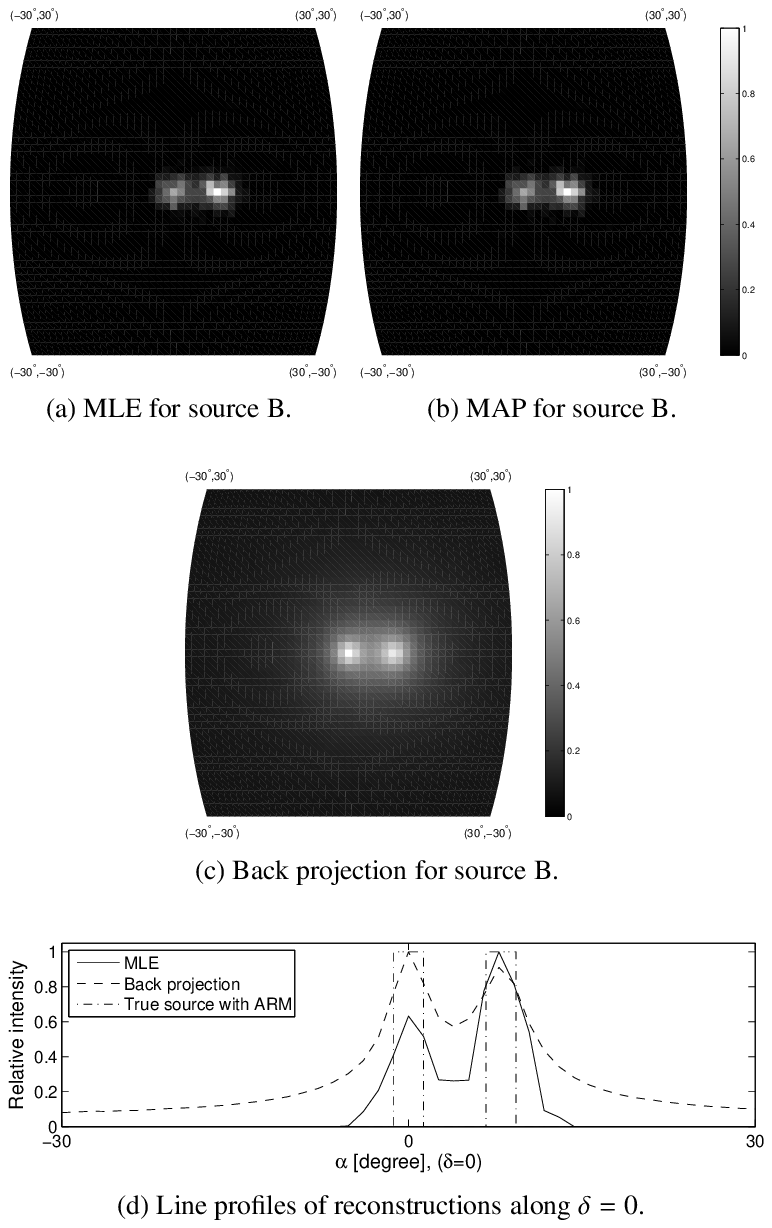}
  \caption{Reconstructed images for two point sources B in 
    figs.\,\ref{fig:sources}b where 147,515 photons were absorbed.
    (a) MLE reconstructions, 
    (b) MAP reconstructions (137 pixels were positive),
    (c) image reconstruction with the back projection, and (d)
    line-profiles of images in (a) and (c) compared with the true
    source profile. Each image was normalized to have $1$ as its
    maximum.}
  \label{fig:result021}
\end{figure}

\begin{figure}[htbp]
  \centering
  \includegraphics[width=1.2\figurewidth]{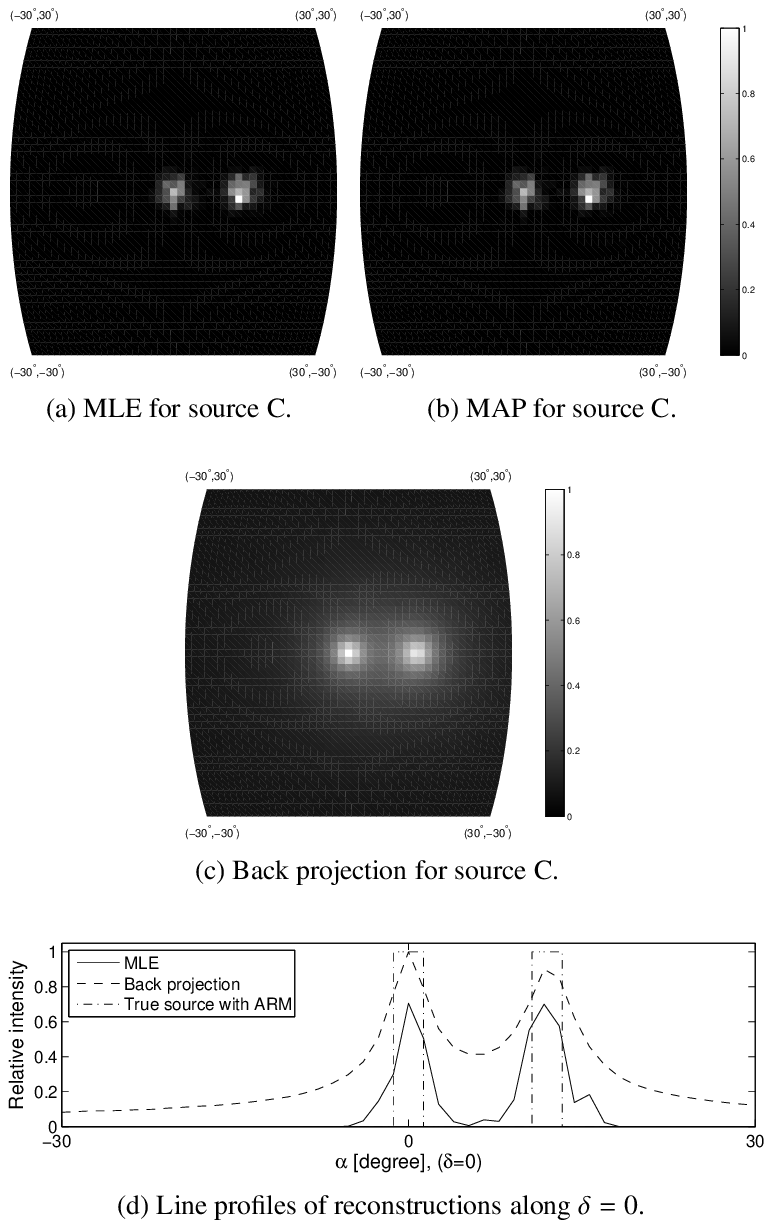}
  \caption{Reconstructed images for two point sources C in 
    figs.\,\ref{fig:sources}c where 150,449 photons were absorbed.
    (a) MLE reconstructions, 
    (b) MAP reconstructions (134 pixels were  positive),
    (c) image reconstruction with the back projection, and (d)
    line-profiles of images in (a) and (c) compared with the true
    source profile. Each image was normalized to have $1$ as its
    maximum.}
  \label{fig:result022}
\end{figure}

Figures \ref{fig:result021} and \ref{fig:result022} show reconstructed
images for sources B and C, respectively. For these sources, the MLE
and the MAP reconstructions separate two points clearly. The line
profiles show that the positive regions of the BME images are similar
to the true sources with the ARM width. 

\begin{figure}[htbp]
  \centering
  \includegraphics[width=1.2\figurewidth]{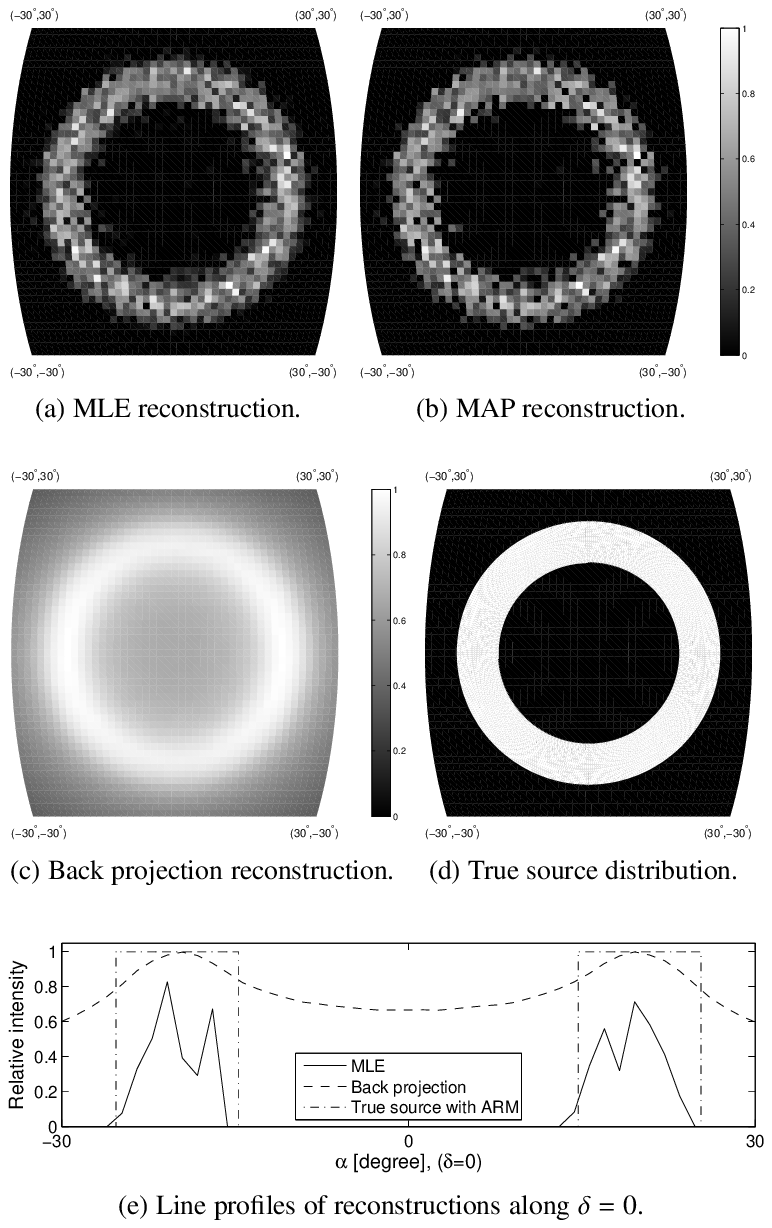}
  \caption{Reconstructed images for source D in
    fig.\,\ref{fig:sources}d where 4,337,897 photons were absorbed.
    (a) MLE reconstruction, 
    (b) MAP reconstruction (668 pixels are positive),
    (c) image reconstruction with the back projection, and (d) true
    distributed gamma-ray source shown comparison.  (e) line-profiles
    of images in (a) and (c) compared with the true source
    profile. Each image was normalized to have $1$ as its maximum }
  \label{fig:resultc}
\end{figure}

Figure \ref{fig:resultc} shows the results for source D. The BME
methods and back-projection clearly differ; the ring-shapes of the
images recovered by the BME methods are clear and similar to the true
distributed source in fig.\,\ref{fig:result0} (d), while the inside of
the ring of the back-projection image is positive. This is clearly
seen from fig.\,\ref{fig:result0} (e). The MAP estimate has only 668
positive pixels, which is around 30\% of the total number of pixels.

All of the BME methods were implemented on a desktop computer (Intel
Core i7-3770S, 3.10\,GHz) with Matlab 2013a (Ubuntu Linux 13.04,
x86-64). Table \ref{tab:comptimes2} summarizes the computational time
and the number of iterations until convergence. The MAP estimates can
be computed faster than the MLE, but the difference becomes smaller
for a distributed source.
\begin{table}[htb]
  \caption{Computational time and iteration numbers of the four
    simulated sources in fig.\,\ref{fig:sources}: the accelerated
    EM algorithm for the MLE and the modified EM algorithm for the MAP.}
  \label{tab:comptimes2}
  \centering
  \begin{tabular}[h]{crrrr}
    \hline
    & 
    \multicolumn{2}{c}{MLE} & 
    \multicolumn{2}{c}{MAP} \\
    {\small source} & 
    {\footnotesize time [sec]} & {\footnotesize iterations} & 
    {\footnotesize time [sec]} & {\footnotesize iterations}\\
    \hline
    A & $3.2\times 10^2$ & 770 &  27 & 1642 \\\hline
    B & $2.6\times 10^2$ & 586 & 40 & 2103 \\\hline
    C & $1.2\times 10^2$ & 274 & 32 & 1554 \\\hline
    D & $2.5\times 10^3$ & 3121 & $1.7\times 10^3$ & 13152 \\
    \hline
  \end{tabular}
\end{table}

\subsection{Discussion}

Numerical results of the image reconstructions were shown in previous
subsections. We provide some discussions in this subsection.

Firstly, we explain why we set the parameter $\beta$ to $0$ for the
MAP estimation. Dirichlet prior with $\beta=1$ corresponds to the MLE,
and the solution becomes sparser as $\beta$ decreases. Since the
Bayesian prior becomes improper for $\beta\le 0$, it is not preferable
to use negative $\beta$. From the experimental results, the MAP
reconstructed images of the MAP estimation with $\beta=0$ were almost
identical to those of the MLE. We did not show any results for
$0<\beta<1$ because the images were almost identical, and the results
were less sparse. Thus, $\beta$ was set to $0$ throughout our
experiments.

Secondly, we consider the convergence speed. Every BME methods was
implemented with an iterative method. Therefore, the convergence speed
is characterized with the computational time of one iteration and the
number of iterations until the convergence.
One iteration of the accelerated EM algorithm takes more time than the
original EM because it uses a line search algorithm. One iteration of
the MAP estimation is computed faster than the original EM. This is
because the MAP estimate is sparse. The E-step shown in
eq.\,\eqref{eq:e-step} is merely a multiplication of a matrix and a
vector, and if the vector is sparse, it can be computed quickly. This
is why the MAP converged faster than the MLE (accelerated EM) even if
the number of iterations was larger than the MLE. However, this effect
becomes less evident when the reconstruction is not sparse. We see
that the difference of the computational time was smaller for source
(D). This is because the final reconstructed image was not sparse.


Thirdly, we discuss the source distribution. We have shown numerical
results for point sources and a ring source. However, it is known that
there is background emission in astronomy. The proposed method can be
applied for the estimation of weak background emission, but a lot of
photons must be collected. This requires a long time for observation
and is not realistic in satellite applications. If the distribution
form of the background is known, for example uniform, and there are
point sources and background emission, it is possible to estimate the
point sources and the level of the background by extending our
framework. However, further details are beyond the scope of this
paper.


Finally, we discuss the quality of the reconstructed images. The BME
reconstructed images were not smooth compared to the
back-projections. Although our results are less smooth, we observed
from the line-profiles that our methods provided good estimates of the
positive regions while the back-projection images became positive for
the whole space. If we know that the true source is a collection of
point sources, it might be possible to estimate their positions as the
peaks of the back-projection image. However, if the source is
distributed, it is difficult to estimate the region by the
back-projection and the proposed BME methods will give better results.

We believe the reason our reconstructions were not smooth partially
comes from the size of the simulation. By collecting more photons for
$p(\V{v}\,|\,\V{u})$ and $s(\V{u})$, we will have smoother
results. This is one of our future works.

\section{Conclusion}
\label{sec:Conclusion}

We have proposed BME methods for Compton camera imaging. Under the
assumption that the distances from the gamma-ray sources are large,
the number of bins can be reduced, making BME methods effective.

We follow the framework of COMPTEL and applied it to a layered
system. Additionally, we propose two extensions of the EM algorithm:
the acceleration of the EM algorithm for the MLE and the modified EM
algorithm for the MAP estimation. Numerical simulations confirm that
the accelerated EM algorithm converges faster than the original EM
algorithm, and the MAP estimate converges quickly into a sparse image.
The proposed methods are promising for astronomy applications. One of
our goal is to apply the propose methods to real astronomy data.

We believe this paper provides a solid framework for the Compton
camera imaging. The EM algorithm for the MLE is a basic approach and
we have shown possible directions for extensions. In the current
paper, we rely on the situation of the astronomy, where the gamma-ray
sources are far from the camera. But if we can relax this assumption,
we may be able to apply similar methods for the measurements of the
distribution of $^{137}\mathrm{Cs}$ in the environment of
Fukushima\cite{jaxa2012} and for medical applications.

\section*{Acknowledgment}

This work was supported by JSPS Grant-in-Aid for Scientific Research
on Innovative Areas, Number 25120008.






\end{document}